\documentstyle[12pt]{article}
\begin{document}
\pagestyle{empty}
\setlength{\oddsidemargin}{0.5cm}
\setlength{\evensidemargin}{0.5cm}
\setlength{\footskip}{2.0cm}
\renewcommand{\thepage}{-- \arabic{page} --}
\vspace*{-2.5cm}
\begin{flushright}
TUM--T31--80/94\\ (hep-ph/9410288)\\ October 1994
\end{flushright}

\vspace*{1.4cm}

\centerline{\large{\bf Studying Structure of Electroweak
Corrections}}

\vspace*{1.7cm}

\renewcommand{\thefootnote}{\arabic{footnote})}
\centerline{\sc \phantom{*)}Zenr\=o HIOKI\,\footnote{Fellow of the
Alexander von Humboldt Foundation}$^,$\footnote{Address after
Nov.$\, 1$, 1994: Institute of Theoretical Physics, University of
Tokushima,\\ \hspace*{0.6cm}Tokushima 770, JAPAN
(E-mail: hioki@ias.tokushima-u.ac.jp)}}

\vspace*{1.7cm}

\centerline{\sl Physik Department,\ Technische Universit\"at\
M\"unchen}

\vskip 0.3cm
\centerline{\sl D-85748 Garching,\ F.$\,$R.$\,$GERMANY}

\vspace*{2.1cm}

\centerline{ABSTRACT}

\vspace*{0.3cm}
\baselineskip=20pt plus 0.1pt minus 0.1pt
Several different effects in electroweak quantum corrections are
explored separately through the latest data on the weak-boson masses.
The leading-log approximation, the improved-Born approximation and
the non-decoupling top-quark effects are studied without depending
on the recent CDF data of $m_t$, and the results are given in a form
independent of the Higgs mass. On the other hand, the bosonic- and
the non-decoupling Higgs effects are examined by fully taking account
of those CDF data. It is emphasized that future precision
measurements of $M_W$ and $m_t$ are considerably significant not only
for further studies of the electroweak theory at higher confidence
level but also for new physics searches beyond it.

\vfill
\newpage
\renewcommand{\thefootnote}{\sharp\arabic{footnote}}
\renewcommand{\theequation}{\arabic{section}.\arabic{equation}}
\pagestyle{plain}
\baselineskip=21.0pt plus 0.2pt minus 0.1pt
\setcounter{footnote}{0}

\section*{\normalsize\bf$\mbox{\boldmath $\S$}\!\!$
1. Introduction}

\vspace*{-0.3cm}
Analyses of electroweak radiative corrections have been extensively
carried out since high-energy experiments at $O(10^2)$ GeV scale
started at SLAC, FNAL and CERN. In particular, a lot of data on
$Z$ from LEP collaborations at CERN have enabled us to test the
minimal standard electroweak theory (hereafter simply ``electroweak
theory'') with considerable accuracy \cite{Pretop}-\cite{NOV94} (see
also \cite{LR} as the latest review articles). Furthermore, CDF
collaboration at FNAL Tevatron collider reported recently some
evidence on top-quark pair productions and estimated thereby its
mass as $m_t^{exp}=174\pm 17$ GeV \cite{TOP}, while it was known in
the precision analyses {\it before this report} that $m_t$ needs to
be 160-180 GeV to describe all the data in the framework of the
electroweak theory.

We surely have come to a position to study ``fine structure'' of
EW (electroweak) corrections as a next step, i.e., to study various
parts of them separately which have different properties. In fact,
several papers have so far appeared for such a purpose: I examined
first whether it is enough for describing $W/Z$-mass data to take
into account only the leading-log terms, and found some indication
for other-type corrections \cite{ZH90}. Then I proposed an effective
procedure to study the non-decoupling top-quark corrections, which
are the main part of such ``non-leading-log'' type contributions
\cite{ZH92}. Novikov et al. studied in \cite{NOV93} the Born
approximation based on $\alpha(M_Z)$ instead of $\alpha$ (so-called
``improved-Born'' approximation), and found that it could reproduce
all electroweak precision data up to 1993 within the $1\sigma$
accuracy (see also \cite{BKKS}). Sirlin observed, however, non-Born
effects in low-energy hadron decays \cite{Sir}, while Novikov et al.
and I also found some deviation from this approximation in 1994
high-energy data \cite{NOV94,ZH94}. From a slightly different point
of view, Dittmaier et al. \cite{DSKK}, and Gambino and Sirlin
\cite{GS} examined lately the bosonic part of the corrections by
using LEP, SLC and Tevatron data, and pointed out their significance.

In this article, I would like to carry out further studies on these
subjects using the latest data on the weak-boson masses. There is no
room for an objection on using $M_Z^{exp}(=91.1888\pm 0.0044$ GeV
\cite{Sch}), while the reason why I focus on the $W$-mass among
others is as follows: All the other high-energy precision data are
those on $Z$ boson or those at $\sqrt{s}\simeq M_Z$. Their accuracy
is now reaching the expected highest level. On the contrary, we can
expect much more precise determination of $M_W$ at Tevatron and LEP
II in the near future, although the present $M_W^{exp}$ by UA2+CDF+D0
has already a fairly good precision. For comparison, the present LEP
data on the $Z$-width are combined to give ${\mit\Gamma}_Z^{exp}=
2.4974\pm 0.0038$ GeV \cite{Sch}, i.e., $\pm$0.15 \% precision. On
the other hand, $M_W^{exp}$ is $80.23\pm 0.18$ GeV ($\pm$0.22 \%
precision) \cite{DFHKS}, i.e., already comparable to
${\mit\Gamma}_Z^{exp}$, and its precision reaches $\pm$0.06 \% once
$M_W$ is measured with an error of $\pm$50 MeV at LEP II
\cite{Kaw}. Moreover, the weak-boson mass relation derived from the
radiative corrections to $G_F$ (the $M_W$-$M_Z$ relation) is the
freest quantity from various strong-interaction effects, and the
meaning of the $W/Z$ masses is of course quite clear.\footnote{
    Another popular quantity in electroweak analyses is
    $\sin^2\theta_W$, but it has a number of different definitions in
    contrast to the weak-boson masses. (Langacker calls this
    situation ``great confusion'' in \cite{LR}.)}\ 
Therefore, we can expect very clean tests through the weak-boson mass
measurements.

Although the CDF results on the top quark is quite important,
its final establishment must come after the confirmation by D0
collaboration. Therefore, I will divide the analyses into two
parts:\\
In $\S\,$3, $m_t$-independent analyses are carried out.\footnote{This
    does not mean that the analyses are completely free from the CDF
    data. I will use the data as a peace of experimental information
    on $m_t$, the meaning of which will become clear at the end of
    $\S\,$2.}\ 
What I study there are the leading-logarithmic approximation, the
improved-Born approximation, and the non-decoupling top-quark
effects. Testing the last one is particularly important because
the existence of such effects is a characteristic feature of theories
in which particle masses are produced through spontaneous symmetry
breakdown plus large Yukawa couplings.\\
In $\S\,$4, on the other hand, I will study other-type corrections by
fully using the CDF data on $m_t$. The EW corrections studied in
$\S\,$3 are essentially those from fermion loops. Theoretically,
however, those from $W$, $Z$ and the Higgs (i.e., bosonic
contributions) are also important. If the top quark is actually very
heavy as the CDF data show, we have a good chance to detect the
bosonic contribution. This is because the fermionic leading-log
terms and the non-decoupling top-quark terms work to cancel each
other, and consequently the role of the non-fermionic corrections
becomes relatively more significant.

Prior to these two sections, a brief review of the EW corrections to
the weak-boson masses is given in $\S\,$2. The final section is
devoted for a conclusion and discussions, where some negative
indication to the electroweak theory is pointed out, though it is
never serious at present. Since a lot of papers have so far appeared
in which comprehensive analyses are performed, as mentioned
in the beginning. I wish, therefore, to show here a little different
aspect of the precision analyses, which is complementary to those
comprehensive analyses.

\vspace*{0.3cm}
\section*{\normalsize\bf$\mbox{\boldmath $\S$}\!\!$
2. EW Corrections to
$\mbox{\boldmath $W$}\!$/$\mbox{\boldmath $Z$}$-masses}
\setcounter{section}{2}\setcounter{equation}{0}

\vspace*{-0.3cm}
I start this section with instructive remarks. Suppose we are trying
to test in a theory the existence of some effects phenomenologically.
Then, we have to show that the following two conditions are
simultaneously satisfied:
\begin{itemize}
\item The theory cannot reproduce the data without the terms under
   consideration, no matter how we vary the remaining free
   parameters.
\item The theory can be consistent with the data by adjusting the
   free parameters {\it appropriately} (i.e., within experimentally
   and theoretically allowed range), once the corresponding terms are
   taken into account.
\end{itemize}
Needless to say, we have to have data and theoretical calculations
precise enough to distinguish these two clearly. In those analyses
it is safer to be conservative: That is, when we check the first
criterion, the smaller number of data we rely on, the more certain
the result is. On the contrary, for checking the second criterion, it
is most trustworthy if we can get a definite conclusion after taking
into account all the existing data, preliminary or not.

Now, let us proceed to the main theme. As I mentioned in $\S\,$1, the
tool of my analyses is the electroweak corrections to the $W/Z$
masses. Through the $O(\alpha)$ corrections to the muon-decay
amplitude, these masses are connected as
\begin{eqnarray}
M_W^2={1\over 2}M_Z^2
\biggl\{ 1+
\sqrt{\smash{1-{{2\sqrt{2}\pi\alpha}
\over{M_Z^2 G_F (1-{\mit\Delta}r)}}}
\vphantom{A^2\over A}
}~\biggr\}. \label{eq21}
\end{eqnarray}
Here ${\mit\Delta}r$ expresses the corrections, and it consists of
several terms with different properties:
\begin{eqnarray}
{\mit\Delta}r={\mit\Delta}r[\ell.\ell.]+{\mit\Delta}r[m_t]+
{\mit\Delta}r[m_{\phi}]+{\mit\Delta}r[\alpha]. \label{eq22}
\end{eqnarray}
${\mit\Delta}r[\ell.\ell.]$ is the leading-log terms from the light
charged fermions, ${\mit\Delta}r[m_t]$ and ${\mit\Delta}r[m_{\phi}]$
express the non-decoupling top-quark and Higgs-boson effects
respectively, and ${\mit\Delta}r[\alpha]$ is the remaining
$O(\alpha)$ non-leading terms.\footnote{The explicit forms of the
    first three terms are given in $\S\,$3 or $\S\,$4, while the
    last one is too lengthy to show in this article. See, e.g., the
    Appendix of the second paper in \cite{ZH90}.}\ 

Equation (\ref{eq21}) is therefore a formula based on the one-loop
calculations (with resummation of the leading-log terms by the
replacement $(1+{\mit\Delta}r)$ $\rightarrow$ $1/(1-{\mit\Delta}r)$).
Over the past several years, some corrections beyond the one-loop
approximation have been computed to it. They are two-loop top-quark
corrections \cite{BBCCV} and QCD corrections up to
$O(\alpha_{\rm QCD}^2)$ \cite{HKl} for ${\mit\Delta}r[m_t]$. As a
result, we have now a formula including $O(\alpha\alpha_{\rm QCD}^2)$
and $O(\alpha^2 m_t^4)$ effects (see also \cite{FKS} as reviews). In
the following, $M_W$ is always computed by incorporating all of these
higher-order terms as well, although I will express the whole
corrections with these terms also as ${\mit\Delta}r$ for simplicity.

Let us see here what we can say about the whole radiative
corrections as a simple example of applications of the $M_W$-$M_Z$
relation and the two criterions I explained in the beginning. The
$W$-mass is computed thereby as
\begin{eqnarray}
M_W^{(0)}=80.941\pm 0.005\ {\rm GeV\ \ and}\ \
M_W=80.33\pm 0.11\ {\rm GeV} \label{eq23}
\end{eqnarray}
for $M_Z^{exp}=91.1888\pm 0.0044$ GeV \cite{Sch} (and
$G_F^{exp}$=$1.16639\times 10^{-5}$ GeV$^{-2}$), where $M_W^{(0)}$
and $M_W$ are those without and with the corrections respectively,
and $M_W$ is for $m_t^{exp}=174 \pm 17$ GeV \cite{TOP},
$m_{\phi}=300$ GeV and $\alpha_{\rm QCD}(M_Z)$=0.118.

As is easily found from Eq.(\ref{eq21}), $M_W^{(0)}$ never depends
on $m_{t,\phi}$. So, we can conclude from $M_W^{(0)}-M_W^{exp}=
0.71\pm 0.18$ GeV and $M_W-M_W^{exp}=0.10\pm 0.21$ GeV that
\begin{itemize}
\item $M_W^{(0)}$ is in disagreement with $M_W^{exp}$ at more than
3.9$\sigma$ (99.99 \% C.L.),
\item $M_W$ can be consistent with the data for, e.g., $m_{\phi}=300$
GeV, which is allowed by the present data $m_{\phi}>61.5$ GeV
\cite{Kob}.
\end{itemize}
That is, the two criterions are both clearly satisfied, by which
the existence of radiative corrections is confirmed. Radiative
corrections were established at $3\sigma$ level already in the
analyses in \cite{Ama}, but where one had to fully use all the
available low- and high-energy data. We can now achieve the same
accuracy via the weak-boson masses alone. Analyses in the following
sections are performed in the same way as this, so I do not repeat
the explanation on the second criterion below since it is common to
all analyses.

Finally, I wish to explain the titles of $\S\,$3 and $\S\,$4. I mean
by ``$m_t$ independent'' that we can test without using $m_t^{exp}$
whether the first criterion is satisfied or not. For the second
criterion, we should use all available data to constrain parameter
space, as already mentioned.

\vspace*{0.3cm}
\section*{\normalsize\bf$\mbox{\boldmath $\S$}\!\!$
3. Top-mass Independent Analyses}
\setcounter{section}{3}\setcounter{equation}{0}

\vspace*{-0.3cm}
Let me first test the validity of the leading-log ($\ell.\ell.$)
approximation. As is well-known, leading-log corrections can be
easily computed via the one-loop renormalization group equations
\cite{Mai}. It
means that any detailed loop calculations are unnecessary if
we can describe data within the $\ell.\ell.$ approximation.

The explicit form of ${\mit\Delta}r[\ell.\ell.]$ in Eq.(\ref{eq22})
is given by
\begin{eqnarray}
{\mit\Delta}r[\ell.\ell.]=-\frac{2\alpha}{3\pi}\sum_{f(\neq t)}Q_f^2
\ln\Bigl(\frac{m_f}{M_Z}\Bigr),
\label{eq31}
\end{eqnarray}
where $Q_f$ is the electric charge of fermion $f$ in the
proton-charge unit, and the sum is both on the flavors and colors
except for the top quark. Since
$\alpha/(1-{\mit\Delta}r[\ell.\ell.])$ is the QED running coupling at
$M_Z$ scale within this approximation, I express this combination as
$\alpha^{(\ell.\ell.)}(M_Z)$. According to Jegerlehner's computations
\cite{Jeg}, $\alpha^{(\ell.\ell.)}(M_Z)$ is estimated to be
$\alpha^{(\ell.\ell.)}(M_Z)=1/(127.69\pm 0.12)$, and thereby the
$W$-mass is calculated through
\begin{eqnarray}
M_W^2[\ell.\ell.]={1\over 2}M_Z^2
\biggl\{ 1+
\sqrt{\smash{1-{{2\sqrt{2}\pi\alpha^{(\ell.\ell.)}(M_Z)}
\over{M_Z^2 G_F}}}
\vphantom{A^2\over A}
}~\biggr\} \label{eq32}
\end{eqnarray}
as $M_W[\ell.\ell.]=79.798\pm 0.017$ GeV. From this and
$M_W^{exp}=80.23\pm 0.18$ GeV \cite{DFHKS}, we get
\begin{eqnarray}
M_W^{exp}-M_W[\ell.\ell.]~=~0.43\pm 0.18~{\rm GeV}. \label{eq33}
\end{eqnarray}
Therefore, they are in disagreement with each other at about $2.4
\sigma$ (98.4 \% C.L.), which means some non-$\ell.\ell.$ corrections
are required at this level. The precision level has now become much
higher than in my previous analysis \cite{ZH90}, where the
disagreement level was $1.3\sigma$ (80.6 \% C.L.).

This $\ell.\ell.$ approximation can be improved by taking account of
some finite terms which are derived by calculating the fermionic
one-loop corrections to the photon self-energy explicitly. Indeed,
Novikov et al. \cite{NOV93} found that all the precision data up to
1993 are reproduced at $1\sigma$ level by using
\begin{eqnarray}
\alpha(M_Z)=
\frac\alpha{\displaystyle{1+\frac{2\alpha}{3\pi}\sum_{f(\neq t)}
Q_f^2\Bigl\{\ln\Bigl(\frac{m_f}{M_Z}\Bigr)+\frac 56\Bigr\}}}
\label{eq34}
\end{eqnarray}
instead of $\alpha^{(\ell.\ell.)}(M_Z)$, where $\alpha(M_Z)$ is known
to be $1/(128.87\pm 0.12)$ \cite{Jeg}.
I examine next whether this ``Improved-Born'' approximation still
works or not. The $W$-mass is calculated within this approximation as
\begin{eqnarray}
M_W[{\rm Born}]=79.957\pm 0.017~{\rm GeV},
\label{eq35}
\end{eqnarray}
which leads to
\begin{eqnarray}
M_W^{exp}-M_W[{\rm Born}]~=~0.27\pm 0.18~{\rm GeV}. \label{eq36}
\end{eqnarray}
This means that the Improved-Born approximation is also in
disagreement with the data now at $1.5\sigma$, which corresponds to
about 86.6 \%\ C.L.. Although the precision is not yet sufficiently
high, it indicates some non-Born terms are needed which give a
positive contribution to the $W$-mass. It is noteworthy since the
electroweak theory predicts such positive non-Born type corrections
unless the Higgs is extremely heavy (beyond TeV scale).

As the final work in this section, I study the non-decoupling
top-quark contribution. The significance of testing it was already
stressed in $\S\,$1.
The non-decoupling top contribution to ${\mit\Delta}r$ is
\begin{eqnarray}
&&{\mit\Delta}r[m_t]=-{\alpha\over{16\pi{\sl s}_W^2}}
\biggl\{ {3\over{{\sl s}_W^2 M_Z^2}}m_t^2
+4\biggl({{\sl c}_W^2\over{\sl s}_W^2}-{1\over 3}
-{{3m_b^2}\over{{\sl s}_W^2 M_Z^2}}\biggr)
\ln\Bigl({m_t\over M_Z}\Bigr)\biggr\} \nonumber\\
&&\phantom{{\mit\Delta}r[m_t]}
\vphantom{{{3m_b^2}\over{{\sl s}_W^2 M_Z^2}}}
+\ [\ higher\ order\ corrections\ ],
\label{eq37}
\end{eqnarray}
where ${\sl c}_W^{\phantom 2}\equiv M_W/M_Z$ and ${\sl s}_W^2=1-
{\sl c}_W^2$. Let me briefly summarize my previous work for this
test. What I proposed in \cite{ZH90} is to study what will happen if
${\mit\Delta}r[m_t]$ would not exist, i.e., to compute the $W$-mass
by using the following ${\mit\Delta}r'$ instead of ${\mit\Delta}r$ in
Eq.(\ref{eq21}):
\begin{eqnarray}
{\mit\Delta}r'\equiv {\mit\Delta}r-{\mit\Delta}r[m_t].  \label{eq38}
\end{eqnarray}
The resultant $W$-mass is denoted as $M_W'$. The important point is
to subtract not only $m_t^2$ term but also $\ln(m_t/M_Z)$ term,
though the latter produces only very small effects unless $m_t$ is
extremely large. ${\mit\Delta}r'$ still includes $m_t$-dependent
terms, but no longer diverges for $m_t\to +\infty$ thanks to this
subtraction. I found that $M_W'$ takes the maximum for the largest
$m_t$ (i.e., $m_t\to+\infty$) and for the smallest $m_{\phi}$ (i.e.,
$m_{\phi}=$61.5 GeV). The accompanying uncertainty was estimated to
be at most 0.03 GeV. Therefore,
\begin{eqnarray}
M_W'\ <\ 79.865\ (\pm 0.030)\ \ {\rm GeV}  \label{eq39}
\end{eqnarray}
holds for any experimentally allowed values of $m_t$ and
$m_{\phi}$.

Comparing this inequality with $M_W^{exp}=80.23\pm 0.18$
GeV, we have
\begin{eqnarray}
M_W^{exp}-M_W'\ >\ 0.36\pm 0.18\ {\rm GeV},
\end{eqnarray}
which shows that $M_W'$ is in disagreement with $M_W^{exp}$ at least
at $2\sigma$ (=95.5 \% C.L.).
This means that the electroweak theory is not able to be
consistent with $M_W^{exp}$ \underline{whatever values $m_t$ and
$m_{\phi}$ take} if the non-decoupling top-quark corrections
${\mit\Delta}r[m_t]$ would not exist. That is,
the latest experimental data of $M_{W,Z}$ demand, \underline{
independent of $m_{\phi}$}, the existence of the non-decoupling
top-quark corrections. It is a very important test of the electroweak
theory as a renormalizable quantum field theory with spontaneous
symmetry breakdown.


\vspace*{0.3cm}
\section*{\normalsize\bf$\mbox{\boldmath $\S$}\!\!$
4. Top-mass Dependent Analyses}
\setcounter{section}{4}\setcounter{equation}{0}

\vspace*{-0.3cm}
As already mentioned in the Introduction, the existence of the
top quark will be established if D0 collaboration also observes
events which show its productions. Indeed, it is quite unlikely that
the top quark does not exist, apart from how heavy it is. Therefore,
I study in this section what we can say when $m_t^{exp}=174\pm 17$
GeV is fully used as an input. Concretely, I wish to examine
non-fermionic contributions to ${\mit\Delta}r$ (i.e., the Higgs and
gauge-boson contributions within the electroweak theory). It has been
pointed out in \cite{DSKK,GS} by using various high-energy data that
such bosonic electroweak corrections are now inevitable. For example,
Gambino and Sirlin applied the same technique as what I proposed in
\cite{ZH92} to $\sin^2\theta_W$ in $\overline{MS}$ scheme, and found
a strong evidence for those effects. I study here whether we can
observe a similar evidence in the $M_W$-$M_Z$ relation.

According to my procedure, I am to show what occurs if the bosonic
contribution would not exist. In this case, in addition to
${\mit\Delta}r[m_{\phi}]$ and the bosonic part of
${\mit\Delta}r[\alpha]$ in Eq.(\ref{eq22}), the two-loop top-quark
corrections have also $m_{\phi}$-dependence, so we have to subtract
them as well. I express the resultant $W$-mass calculated for
${\mit\Delta}r-{\mit\Delta}r[{\rm boson}]$ instead of ${\mit\Delta}r$
as $M_W[{\rm f}]$ since it receives only the fermionic contribution.
$M_W[{\rm f}]$ is computed for $m_t^{exp}=174\pm 17$ GeV as
\begin{eqnarray}
M_W[{\rm f}]=80.44\pm 0.11\ {\rm GeV}. \label{eq41}
\end{eqnarray}
This value is of course independent of the Higgs mass. Equation
(\ref{eq41}) means
\begin{eqnarray}
M_W[{\rm f}]-M_W^{exp}=0.21\pm 0.21\ {\rm GeV}, \label{eq42}
\end{eqnarray}
which shows that some non-fermionic contribution is necessary at
$1\sigma$ level.

It is of course too early to say from this, e.g., that the bosonic
effects were confirmed. Nevertheless, this is an interesting result
since we could observe nothing before: For example, the best
information on $m_t$ before the CDF report was the bound $m_t >$ 131
GeV by D0 \cite{D0}, but we can thereby get only $M_W[{\rm f}] >$
80.19 ($\pm$0.03) GeV (i.e., $M_W[{\rm f}]-M_W^{exp} > -0.04\pm 0.18$
GeV). We will be allowed therefore to conclude that ``the bosonic
effects are starting to appear in the $M_W$-$M_Z$ relation''. All the
results obtained so far are visually represented in the Figure.

\centerline{\bf ------------------------}

\centerline{\bf Figure}

\centerline{\bf ------------------------}

For comparison, let us make the same computation for
${\mit\Delta}M_W^{exp}=\pm 0.05$ GeV and
${\mit\Delta}m_t^{exp}=\pm 5$ GeV, which will be eventually realized
in the future at Tevatron and LEP II. Concretely,
${\mit\Delta}m_t^{exp}=\pm 5$ GeV produces an error of $\pm 0.03$ GeV
in the $W$-mass calculation. Combining this with the
theoretical ambiguity ${\mit\Delta}M_W=\pm 0.03$ GeV, we can compute
$M_W[\cdots]-M_W^{exp}$ with an error of about $\pm 0.07$ GeV. Then,
$M_W[{\rm f}]-M_W^{exp}$ becomes $0.21\pm 0.07$ GeV if the central
value of $M_W^{exp}$ is the same, by which we can confirm the above
statement at $3\sigma$ level.

We can similar way study the QCD corrections (though it does not
match the title of this paper). In this case, I could not find any
inequality like Eq.(\ref{eq39}) which leads to a definite
($m_{\phi}$-independent) statement. We can still get, however, some
information about the role of the QCD corrections: If we remove these
corrections and compute the $W$-mass, we get $M_W$[pure-EW]=80.48
$(\pm 0.11)$ GeV for $m_{\phi}=100$ GeV. This leads to
$M_W$[pure-EW]$-M_W^{exp}=0.25\pm 0.21$ GeV, which means such a
light Higgs is not favored. Indeed, the central value of $M_W^{exp}$
(=80.23 GeV) becomes to require an extremely heavy Higgs: $m_{\phi}>$
3 TeV.\footnote{Do not take this value too seriously. Perturbation
    theory is no longer reliable if the Higgs is so heavy \cite{BV}.}
    \ 
This result tells us that we can no longer neglect QCD effects even
in the $M_W$-$M_Z$ relation, the least QCD-dependent quantity.
Anyway, the electroweak theory is saved by the QCD corrections.

It must be very interesting if we can find moreover the existence of
the non-decoupling Higgs effects
\begin{eqnarray}
{\mit\Delta}r[m_{\phi}]={{11\alpha}\over{24\pi{\sl s}_W^2}}
\ln\Bigl({m_{\phi}\over M_Z}\Bigr),  \label{eq43}
\end{eqnarray}
since we still have no phenomenological indication for the Higgs
boson. Then, can we in fact perform such a test? It depends on how
heavy the Higgs is: If it is much heavier than the weak bosons, then
we may be able to test it. If not, however, that test will lose its
meaning essentially, since ${\mit\Delta}r[m_{\phi}]$ comes from the
expansion of terms like $\int^1_0 dx\ln\{m^2_{\phi}(1-x)+M^2_Z
x-M^2_Z x(1-x)\}$ in powers of $M_Z/m_{\phi}$. Here, let us simply
assume as an example that we have gained in some way (e.g., at LHC) a
bound $m_{\phi} > 500$ GeV. Then, for ${\mit\Delta}r''\equiv
{\mit\Delta}r-{\mit\Delta}r[m_{\phi}]$, the $W$-mass (written as
$M_W''$) satisfies $M_W'' > 80.46\pm 0.11$ GeV, where the
non-decoupling $m_{\phi}$ terms in the two-loop top-quark corrections
were also eliminated. This inequality leads us to $M_W'' -M^{exp}_W >
0.23\pm 0.21$ GeV.

It seems therefore that we may have a chance to get an indirect
evidence of the Higgs boson even if future accelerators would fail to
discover it. Indeed, the central value of the present $M_W^{exp}$
needs a very heavy Higgs (about 1 TeV \cite{HN}: see the next
section) though its still large uncertainty allows also a light
Higgs. Fortunately or unfortunately, however, the present LEP+SLC
data require a light Higgs
($\lower0.5ex\hbox{$\buildrel <\over\sim$}$
200-300 GeV) \cite{Pretop}. So, if we get any evidence that the Higgs
is heavier than 500 GeV, it means that the electroweak theory falls
into a trouble. In such a case, it will have little meaning to study
various radiative corrections within only the minimal electroweak
theory. I will discuss this problem briefly in the next section.

\vspace*{0.3cm}
\section*{\normalsize\bf$\mbox{\boldmath $\S$}\!\!$
5. Conclusion and Discussions}
\setcounter{section}{5}\setcounter{equation}{0}

\vspace*{-0.3cm}
I have carried out here analyses on (1) $m_t$-independent and (2)
$m_t$-dependent electroweak quantum corrections in the weak-boson
mass relation.

In the former part, I tested the leading-log approximation, the
Improved-Born approximation, and also the non-decoupling top
corrections. We could thereby conclude that non-$\ell.\ell.$ type
corrections, non-Born type corrections, and non-decoupling $m_t$
contribution are required respectively at about $2.4\sigma$,
$1.5\sigma$, and $2.0\sigma$ level by the recent data on $M_{W,Z}$.
This is a clean, though not yet perfect, test of non-trivial
corrections which has the least dependence on hadronic contributions.

Concerning the latter part, we could observe a small indication for
non-fermionic contributions (at $1\sigma$ level, though). This comes
from large cancellation between the light-fermion and heavy-top
terms, and such contributions can be interpreted as those from the
bosonic ($W/Z$ and the Higgs) corrections. Furthermore, it seemed to
be possible to test the non-decoupling Higgs effects if the Higgs
boson is heavy (e.g., $\lower0.5ex\hbox{$\buildrel >\over\sim$}$ 500
GeV).

On the $m_t$-dependent corrections, however, supplementary
discussions are necessary as mentioned in the end of the previous
section. In $\S\,$2, the $W$-mass with the whole corrections for
$m_t^{exp}=174\pm 17$ GeV was computed for $m_{\phi}=300$ GeV.
However, in order for $M_W|_{m_t=174\ {\rm GeV}}$ to reproduce the
central value of $M_W^{exp}$ (80.23 GeV) within the electroweak
theory, the Higgs mass needs to be 1.1-1.2 TeV \cite{HN}. Even if we
limit discussions to perturbation calculations, such an extremely
heavy Higgs will cause several problems \cite{DM,DKR}. Moreover, the
present LEP and SLC data require a light Higgs boson:
$m_{\phi}$ $\lower0.5ex\hbox{$\buildrel <\over\sim$}$ 200-300 GeV
though at $1\sigma$ level \cite{Pretop}. This means that we might
be caught in a kind of dilemma.

At present, it is never serious since $m_{\phi}$ as low as 60 GeV is
also allowed if we take into account
${\mit\Delta}m_t^{exp}=\pm 17$ GeV and ${\mit\Delta}M_W^{exp}
=\pm 0.18$ GeV ($M_W-M_W^{exp}=0.20\pm 0.21$ GeV for $m_{\phi}=60$
GeV). That is why $\chi^2$ takes its minimum at low $m_{\phi}$
even when $M_W^{exp}$ is taken into account in an analysis. However,
if ${\mit\Delta}M_W^{exp}=\pm 0.05$ GeV and ${\mit\Delta}m_t^{exp}=
\pm 5$ GeV are realized as I assumed in $\S\,$4, a constraint from
the $W$-mass becomes stronger. As an example, let us assume that the
central values of $M_W^{exp}$ and $m_t^{exp}$ do not change. Then,
$M_W-M_W^{exp}$ becomes $0.10\pm 0.07$ GeV for $m_{\phi}=300$ GeV. It
means that $m_{\phi}=300$ GeV is ruled out at $1.4\sigma$ level
within the minimal standard electroweak theory. Similarly, even
$m_{\phi}=400$ GeV is not favored though at $1.1\sigma$ level
($M_W-M_W^{exp}=0.09\pm 0.07$ GeV).

It is of course premature to say thereby, e.g., that we can see some
new-physics effects in the very near future. In fact, the above
arguments are only an example, and strongly depend on the values
of $M_W^{exp}$ and $m_t^{exp}$. The UA2 and CDF data require a little
heavier $W$-mass while D0 gives a lighter one as
$M_W^{exp}[{\rm UA2}]=80.36\pm 0.37$ GeV, $M_W^{exp}[{\rm CDF}]=80.38
\pm 0.23$ GeV, and $M_W^{exp}[{\rm D0}]=79.86\pm 0.40$ GeV
\cite{DFHKS}. If a future experimental value converges to a higher
value like $M_W^{exp}[{\rm UA2,CDF}]$, the above problem disappears
and we also be able to give the conclusions for $\S\,$3 more strongly
(but the conclusion for $\S\,$4 becomes weaker). On the contrary, if
$M_W^{exp}$ moves to $M_W^{exp}[{\rm D0}]$, the situation becomes
much more serious. In that case, we will have to consider in earnest
new physics which produces opposite contributions to $M_W$ and to the
other quantities. The top-mass measurements can also affect these
arguments similarly: If $m_t$ moves to a lighter value, the
electroweak theory is OK, but if it is found to be heavier, the
problem is again serious.

Precise measurements of $M_W$ and $m_t$ are therefore considerably
significant not only for precision tests of the electroweak theory
but also for new physics searches beyond this theory.

\vspace*{0.6cm}
\centerline{ACKNOWLEDGEMENTS}

\vspace*{0.3cm}
I would like to thank K. Riesselmann and R. Najima for stimulating
discussions, S. Errede for sending me useful information on $M_W$, J.
Fleischer for correspondence on their calculations, and A. J. Buras
for reading the manuscript and valuable comments on it. I am also
grateful to A. J. Buras and all the members of the particle theory
group at TU-M\"unchen for their warm hospitality, and K. Riesselmann,
M. Lautenbacher and S. Herrlich for introducing me to the computer
system of this laboratory.


\newpage
\setlength{\unitlength}{0.1mm}
\vspace*{-1.5cm}
\begin{picture}(1380,1700)(0,-100)
\put(0,100){\line(0,1){1390}}
\put(0,100){\vector(1,0){1350}}
\put(500,100){\line(0,1){1390}}
\multiput(850,100)(0,50){28}{\line(0,1){40}}
\put(600,100){\line(0,1){20}}
\put(1100,100){\line(0,1){20}}
\put(540,50){\makebox(100,50)[bl]{$-0.5$}}
\put(840,50){\makebox(100,50)[bl]{$0$}}
\put(1075,50){\makebox(100,50)[bl]{$0.5$}}
\put(780,1520){\makebox(500,50)[bl]{``happy'' line}}
\put(700,-20){\makebox(500,50)[bl]{$M_W[\cdots]-M_W^{exp}$ (GeV)}}
\put(30,1400){\makebox(400,70)[bl]{$M_W^{(0)}$}}
\put(1205,1400){\circle*{20}}
\put(1115,1400){\line(1,0){180}}
\put(30,1340){\makebox(400,70)[bl]{($=80.941\pm 0.005$ GeV)}}
\put(30,1200){\makebox(400,70)[bl]{$M_W[\ell.\ell.]$}}
\put(635,1200){\circle*{20}}
\put(545,1200){\line(1,0){180}}
\put(30,1140){\makebox(400,70)[bl]{($=79.798\pm 0.017$ GeV)}}
\put(30,1000){\makebox(400,70)[bl]{$M_W[{\rm Born}]$}}
\put(715,1000){\circle*{20}}
\put(625,1000){\line(1,0){180}}
\put(30,940){\makebox(400,70)[bl]{($=79.957\pm 0.017$ GeV)}}
\put(30,800){\makebox(400,70)[bl]{$M_W'$}}
\put(667.5,800){\circle*{20}}
\put(757.5,800){\vector(-1,0){220}}
\put(30,740){\makebox(400,70)[bl]{($<79.865\pm 0.030$ GeV)}}
\put(30,600){\makebox(400,70)[bl]{$M_W[{\rm f}]$}}
\put(955,600){\circle*{20}}
\put(850,600){\line(1,0){210}}
\put(30,540){\makebox(400,70)[bl]{($=80.44\pm 0.11$ GeV)}}
\put(30,360){\makebox(400,70)[bl]{$M_W$}}
\put(30,300){\makebox(400,70)[bl]{($=80.40\pm 0.11$ GeV [{\rm a}])}}
\put(935,310){\circle*{20}}
\put(830,310){\line(1,0){210}}
\put(30,240){\makebox(400,70)[bl]{($=80.33\pm 0.14$ GeV [{\rm b}])}}
\put(900,250){\circle*{20}}
\put(785,250){\line(1,0){230}}
\end{picture}

\vspace*{0.5cm}
\centerline{\bf Figure}

Deviations of $W$-masses calculated in various approximations from
$M_W^{exp}=80.23\pm 0.18$ GeV, where $M_W[{\rm a,b}]$ are those with
the full corrections for $m_{\phi}=100$ GeV and for 61.5 GeV $<\
m_{\phi}\ <$ 1 TeV (the central value is 300 GeV) respectively.
Only $M_W[{\rm a,b}]-M_W^{exp}$ cross the ``happy'' line, and
$M_W[{\rm f}]-M_W^{exp}$ barely touches the line.
\end{document}